\documentstyle[amssymb,aps,preprint,prl]{revtex}
\draft

\begin{document}
\title{Instabilities and disorder-driven first-order transition of the vortex lattice}
\author{Y. Paltiel$^1$, E. Zeldov$^1$, Y. Myasoedov$^1$, M. L. Rappaport$^1$,
G. Jung$^{1,2}$, S. Bhattacharya$^{3,4}$, M. J. Higgins$^3$, Z. L. Xiao$^5$, E. Y. Andrei$^5$, P. L. Gammel$^6$, and D. J. Bishop$^6$} 
\address{$^1$Department of Condensed Matter Physics, Weizmann Institute of
Science, Rehovot 76100, Israel}
\address{$^2$Department of Physics, Ben-Gurion University of the Negev, Beer-Sheva 84105, Israel}
\address{$^3$NEC Research Institute, 4 Independence Way, Princeton, New Jersey 08540} 
\address{$^4$Tata Institute of Fundamental Research, Mumbai-400005, India}
\address{$^5$Department of Physics and Astronomy, Rutgers University, Piscataway,
New Jersey 08855}
\address{$^6$Bell Laboratories, Lucent Technologies, Murray Hill, New Jersey 07974}

\date{\today}
\maketitle

\begin{abstract}
Transport studies in a Corbino disk geometry suggest that the Bragg
glass phase undergoes a first-order transition into a disordered solid. This transition shows a sharp reentrant behavior at low fields. In contrast, in the conventional strip configuration, the phase transition is obscured by the injection of the disordered vortices through the sample edges, which results in the commonly observed vortex instabilities and smearing of the peak effect in NbSe$_2$ crystals. These features are found to be absent in the Corbino geometry, in which the circulating vortices do not cross the sample edges.
\end{abstract}

 

\newpage

The nature of the disorder-driven solid-solid transition in the vortex matter in superconductors \cite{gl,kier,en,gingras,larkin,anis}, and the associated instabilities \cite{hend98,welp,yosna,marley95,kwok,danna,shobo95,hend96,baner98,baner99,pardo,yaron,won,degroot,giller,van,gordeev,zhukov}, have recently attracted wide attention. A number of anomalous instability phenomena were reported, which include memory effects \cite{hend98}, frequency and bias dependence \cite{hend98,welp,yosna}, low frequency noise \cite{marley95,kwok,danna}, history dependent response \cite{shobo95,hend96,baner98,baner99,pardo,yaron,won,degroot,giller,van}, slow voltage oscillations \cite{kwok,gordeev},  and negative dynamic creep \cite{zhukov}.
NbSe$_2$ is a very convenient system for studying these phenomena since it displays a pronounced peak effect (PE) in the critical current $I_c$ below the upper critical field $H_{c2}$. The PE constitutes a transformation of the quasi-ordered Bragg glass phase \cite{gl,kier,en,gingras} below the PE, to a highly disordered phase in the PE region \cite{gammel98}. Although several sharp features have been observed at the PE \cite{baner98,baner99,won} and various models for the PE suggested \cite{larkin,won}, there is currently no general consensus regarding the underlying nature of the disordered phase (DP) and of the corresponding order-disorder transition.

 An important property of the disorder-driven transition is that the DP can be `supercooled' to well below the transition by a field-cooling procedure, where it remains {\em metastable} in the absence of driving currents \cite{hend96,baner98,baner99,pardo,yaron,won}. In this Letter we address the question of how the {\em metastable} DP modifies the apparent transport behavior and what are the actual underlying vortex matter properties in the absence of the metastabilities. Our study is motivated by a recent model, according to which the metastable DP can be formed {\em dynamically} by an edge-contamination mechanism \cite{yosna}. In the presence of a driving current, the flowing vortices have to penetrate into the sample through the surface barriers at the edges \cite{bur}.
Since the barrier height is non-uniform due to material imperfections, the vortices are injected predominantly at the weakest points of the barrier, hence destroying the ordered phase (OP) and creating a metastable DP near the edges.
Consequently, the common experimental procedures inadvertently cause a dynamic admixture of the OP with the metastable DP, thus preventing observation of unperturbed OP and of its behavior in the vicinity of the transition. 
We have carried out transport measurements in a Corbino disk geometry (inset to Fig. 1), in which the vortices circulate in the bulk without crossing the edges, and hence the injection of the DP could be prevented. These measurements are compared with the regular strip configuration in the same crystals.
If the instability effects are a bulk phenomenon, no significant difference between the two geometries is expected, whereas if the sample edges do play a dominant role, qualitatively different behavior should be observed. This paper demonstrates that the two geometries yield strikingly different response and, significantly, the instability effects are eliminated in the Corbino geometry. Furthermore, by removing the complicating effects of the edges, the true bulk dynamics in the vicinity of the transition can be probed for the first time. The results suggest that the disorder-driven transition $T_{DT}$ is possibly a thermodynamic first-order phase transition. In addition, it is found that the $T_{DT}$ line displays a sharp reentrant behavior at low fields in a wide temperature range.

The results presented here were obtained on a Fe (200ppm) doped 2H-NbSe$_2$ single crystal 2.2 $\times $ 1.5 $\times $ 0.04 mm$^{3}$ with $T_c$=5.7 K, which display a significantly broader PE as compared to the pure crystals \cite{hend98,baner98}. Similar data were obtained on three additional samples including one pure crystal. The crystals were cut with a very fine wire-saw to ensure uniform edges. Ag$/$Au contacts were evaporated through a mask designed for a comparative study of the Corbino and strip geometries on the same crystal, using the same voltage contacts +V,-V (Fig. 1 inset). When measuring the Corbino, the current is applied to the +C,-C contacts, while for the strip configuration the +S,-S contacts are used. To prevent vortices from crossing the edges it is crucial to enforce a uniform radial current in the Corbino. The current uniformity is limited by the variations in the local contact resistance along the outer ring electrode. In order to improve the current uniformity, the ring electrode is divided into four quadrants, and 1/4 of the current is applied to each quadrant. Without such a careful current balance, some of the vortices are injected through the sample edges resulting in metastability effects similar to the strip configuration. The Corbino and the strip geometries were usually compared at the same applied current. In both geometries the current density is not uniform throughout the sample. A calibration above $T_c$ shows that the average current density between the +V,-V voltage contacts is higher by about 25\%  in the strip configuration, consistent with a geometrical calculation. Therefore, all voltage readings ($V$) of the strip were divided by about 1.25 in order to obtain the same voltage above $T_c$ in the two geometries. Similarly, the $I_c$ values of the strip were multiplied by the same factor. The {\em ac} measurements were limited to frequencies below 1 kHz to ensure that there were no effects of finite skin depth.
The temperature reproducibility between different runs was about 30 mK. 

Figure 1a shows the critical current vs. temperature measurements in the vicinity of the PE at 2 kOe. The striking difference between the two geometries is evident. While the {\em dc} $I_c$ in the strip geometry shows the usual smooth PE, the Corbino $I_c$ displays a very sharp drop at the maximum of the PE. Similarly, Fig. 1b shows the voltage at a constant current at 2.5 kOe, which displays a very sharp drop at $T_{DT}$ in the Corbino, for both {\em dc} and {\em ac} currents. Within our temperature reproducibility the response in the Corbino is practically frequency independent in the entire field and temperature range, as expected in the absence of the instability phenomena. The sharp transition in the Corbino geometry reflects a disorder driven transition $T_{DT}$ between two thermodynamically stable phases: an OP below $T_{DT}$, which is dominated by the elastic energy and is characterized by a low critical current $J_c^{ord}$, and a DP above $T_{DT}$, which is governed by the pinning energy and has a high $J_c^{dis}$.

We now analyze the transport behavior in the strip geometry. Figure 1 shows that the system has a conventional response with no geometry and no frequency dependence above $T_{DT}$. This is where only one stable phase, the DP, is present. In contrast, the region below $T_{DT}$ is where all the anomalous phenomena in the strip configuration are found \cite{hend98,welp,yosna,marley95,kwok,danna,shobo95,hend96,baner98,baner99,pardo,yaron,won,degroot,giller,van,gordeev,zhukov}.
In this region, in the presence of a driving current, a metastable DP is injected through the sample edges instead of the equilibrium OP \cite{yosna,giller}. The driven metastable DP has a finite lifetime, $\tau_r$, that defines a characteristic relaxation length $L_r$=$v\tau_r$, over which the metastable DP anneals into the OP ($v$ is the vortex drift velocity).
As a result, the measured {\em dc} $I_c$ in the strip is a weighted superposition of $J_c^{ord}$ with the metastable $J_c^{dis}$, given by $I_{c}=d \int_0^WJ_c(x)dx \simeq dL_{r}J_{c}^{dis}+d(W-L_{r})J_{c}^{ord}$, where $d$ and $W$ are the sample thickness and width \cite{yosna}.
For $L_{r}\ll W$ we obtain $I_c = dWJ_c^{ord}$, and $I_c = dWJ_c^{dis}$ for $L_{r}\geq W$. The relaxation length $L_r$ is very sensitive to the proximity to the transition. Near $T_{DT}$ the elastic and the pinning energies are comparable, and therefore the metastable DP has a long lifetime, and hence $L_r$ is large. Further below $T_{DT}$ the driven DP becomes progressively unstable, and therefore $L_r$ drops continuously at lower temperatures, resulting in a smooth decrease of $I_c$ in the strip \cite{note1}. Since $L_r$ starts to decrease immediately below $T_{DT}$, the maximum in $I_c$ of the strip coincides with $T_{DT}$, as seen in Fig. 1a. The enhanced $I_c$ in the strip below $T_{DT}$ causes the {\em dc} voltage response to be significantly suppressed and shifted to lower temperatures compared to the Corbino, as seen in Fig. 1b. This voltage has a smoother temperature and field dependence due to the gradual evolution of the effective $I_c$ in the strip.

The edge contamination also results in a very pronounced frequency dependence below $T_{DT}$. An {\em ac} driving current limits the contamination by the DP to narrow regions, of width $x_d^{ac}$, near the edges of the strip, where the vortices penetrate and exit during the {\em ac} cycle \cite{yosna}. Since $x_d^{ac} \leq L_r$, the $I_c$ measured by an {\em ac} current in the strip is significantly reduced relative to the {\em dc} measurement, as shown in Fig. 1a. 
Also, since $x_d^{ac}$ shrinks with frequency, the voltage increases with frequency and is shifted towards the response of the Corbino as shown in Fig. 1b. {\em The combination of the observed differences in the {\em ac} and {\em dc} properties below $T_{DT}$ in the two geometries, as presented in Fig. 1, cannot be attributed to any known bulk vortex mechanism or inhomogeneities, and clearly demonstrates the key role of the sample edges}. 

The location of the disorder-driven transition line $T_{DT}$ or $H_{DT}$ on the $H-T$ phase diagram, derived from the Corbino data, is shown in Fig. 2. Striking reentrant behavior of $H_{DT}$ is observed. 
At elevated fields, the elastic energy decreases with field resulting in the high-field $H_{DT}$ line when it becomes equal to the pinning energy.
The recent theoretical studies of the disorder-driven transition have mainly focused on this high-field $H_{DT}$ transition \cite{gl,kier,en,gingras}. However, the elastic energy also decreases rapidly at low fields where the vortex interactions start to decrease exponentially, which leads to a mirror-like reentrant $H_{DT}$ line \cite{gl,gingras,larkin}. Similar arguments involving thermal fluctuations lead to the well known prediction of reentrant melting \cite{nelson}. Although observations of a reentrant PE were originally interpreted in terms of reentrant melting \cite{ghosh}, subsequent studies attribute this behavior to a reentrant disordering of the lattice \cite{ban}. Our results demonstrate the first unambiguous reentrant phase diagram with very sharp and pronounced transition lines. In principle, by measuring the high-field $H_{DT}$ line and knowing all the microscopic parameters and the details of the disorder, one should be able to predict the location of the reentrant $H_{DT}$ line and compare it with the experiment. Such an extensive theoretical analysis is beyond the scope of this work and will be presented elsewhere. 

Both the high-field and the reentrant $H_{DT}$ lines can be crossed in a single experimental run by a field sweep at a constant temperature, as presented in Fig. 3 for 4.3 K and 4.6 K. The transport data in the Corbino geometry show remarkably sharp resistive drops at both the low-field and high-field $H_{DT}$ points. Note the almost linear field dependence of the resistance in between the two transitions. Such a linear behavior is one of the hallmarks of the weakly pinned OP, as observed in clean NbSe$_2$ crystals \cite{bhatta}.  The strip geometry shows markedly different behavior. Below the upper $H_{DT}$ in Fig. 3a the resistive onset is gradual and is shifted to a lower field of $H_{on}^{s}$ due to the injection of the metastable DP from the edges, similar to the behavior described in Fig. 1b. Surprisingly, we find that also the instability phenomena display a reentrant behavior, closely following the reentrant $H_{DT}$ line. In this case, however, the metastable DP is present at fields {\it above} the reentrant $H_{DT}$, instead of below the transition, as near the high-field $H_{DT}$. As a result, the reentrant $H_{on}^{s}$ of the strip resides above the reentrant $H_{DT}$. Also the {\em ac} voltage of the strip displays reentrant anomalous behavior, as shown by the dashed curve in Fig. 3a. Near the upper $H_{on}^s$ the {\em ac} data is shifted towards higher fields with respect to the {\em dc} strip response, whereas in the reentrant region a mirror-image like displacement to {\it lower} fields is obtained. 

The proximity to the $H_{DT}$ transition is essential to the enhanced lifetime of the metastable DP. Accordingly, the relaxation length $L_r$ diverges upon approaching the high-field $H_{DT}$ from below and the reentrant $H_{DT}$ from above. In the Corbino geometry the entire area within the triangle formed by the $H_{DT}$ line in Fig. 2 represents a stable OP. In the strip case, however, in this triangle, a metastable DP dynamically coexists with the OP. In the belt area between $H_{on}^s$ and $H_{DT}$, $L_r$ is sufficiently large, such that the $V_{dc}$ of the strip in Fig. 3a is immeasurably small at the applied current of 20 mA. Within the triangular area of $H_{on}^s$, marked by a dashed line, significant vortex motion gradually builds up away from the transition. Yet even here $L_r$ remains finite since the full flux-flow vortex velocity of the Corbino is not attained. An interesting case is shown in Fig. 3b where the phase diagram is crossed vertically at 4.6 K, cutting through the tip of the $H_{on}^s$ triangle (see Fig. 2). In the strip geometry no voltage response is observed at 20 mA except a noisy behavior near 2 kOe. This is the characteristic noise associated with the described instability phenomena, as reported previously \cite{marley95}. 

The inset to Fig. 2 shows $I_c$ at 4.2 K as a function of $H$ on crossing the reentrant $H_{DT}$. One may expect to see here a PE similar to the high-field peak effect. The high-field PE originates from the fact that $I_c$ increases upon crossing from OP to DP, but then drops gradually to zero near $H_{c2}$. In the reentrant case, however, this analogy is not complete, since $I_c$ does not go to zero upon approaching $H_{c1}$. Instead, the $I_c$ in the Corbino shows a sharp increase at $H_{DT}$, but then continues to grow as the field is decreased within the reentrant DP, since the diluted vortices remain strongly pinned individually \cite{ban}. In contrast to the Corbino, the strip configuration shows once again a smooth behavior of $I_c$ at $H_{DT}$, like a mirror-image of Fig. 1. {\em Thus, regardless of whether the order to disorder transition occurs upon increasing or decreasing the field, the injection of the metastable DP from the sample edges always occurs on the OP side of the transition in the vicinity of $H_{DT}$}.

The present findings allow us to derive some conclusions regarding the thermodynamic nature of the $H_{DT}$ transition. Figures 1 and 3, as well as the inset to Fig. 2, show extremely sharp resistive transitions at $H_{DT}$ in the Corbino geometry, which are much sharper than the resistive transition at $T_c$. This observation is indicative of the first-order nature of the disorder-driven transition. Obviously, resistivity is not a thermodynamic probe, and further investigations are required. Yet historically, the first strong indications of a possible first-order melting transition came from similarly sharp resistive kinks in HTS crystals \cite{safar}, which were confirmed thermodynamically only later \cite{melt}. Our results thus imply that the disorder-driven destruction of the Bragg glass is possibly of first order, similar to the thermally driven destruction upon melting. Since both the disordered phase and the vortex liquid do not possess any long range order, it is plausible to expect that their transition into an ordered phase should be of first order, involving topological symmetry breaking in both cases. This means that the second peak transition in HTS \cite{anis}, which is of the same nature as the PE in NbSe$_2$, could be of first order as well, thus forming a unified first-order destruction line of the Bragg glass at all temperatures. Recent theoretical considerations seem to support this scenario \cite{kierfeld}.

In summary, we find that the ordered Bragg glass becomes unstable with respect to disorder at both high and low fields, resulting in a reentrant disorder-driven transition line. By using a Corbino geometry, and thus avoiding the contamination from the sample edges, this transition is found to be very sharp and apparently of first order. The vortex instability phenomena are caused by the injection of metastable disorder through the sample edges, and are therefore absent in the Corbino geometry. The instabilities are present in the strip geometry on the Bragg glass side of the transition along both the high-field and the low-field branches of the transition line.

This work was supported by the US-Israel Binational Science Foundation (BSF), by the Israel Science Foundation - Center of Excellence Program, and by the Alhadeff Research Award. EYA acknowledges support by the NSF.

\newpage 

\ FIGURE CAPTIONS

Fig. 1. (a) The critical current $I_c$ vs. temperature at 2 kOe measured by 
a {\em dc} current in the Corbino geometry ($\blacksquare $), and by {\em dc} ($\square $) and 172 Hz {\em ac} ($\bigcirc $) current in the strip. The $I_c$ is defined at a voltage criterion of 0.5 $\mu$V.
The dashed line is a schematic guide to the eye extrapolation of the $I_c^{dis}$. 
(b) Voltage vs. temperature at 2.5 kOe and 20 mA, using {\em dc} ($\bigcirc $) and 765 Hz {\em ac} ($\square $) current in the Corbino, and {\em dc}, 22, 172 and 765 Hz in the strip.
Inset: the electrode configuration allowing measurements in both the Corbino and strip configurations by using +C,-C and +S,-S current contacts, respectively. The outer diameter of the Corbino electrode is 1.1 mm and the distance between the centers of the voltage contacts is 0.15 mm.

Fig. 2. $H-T$ phase diagram showing the disorder-driven phase transition line $H_{DT}$ with reentrant behavior, as determined from the Corbino data. The ordered phase (OP) is present within the triangular region defined by the solid $H_{DT}$ line, and is surrounded by the DP both at high and low fields. The instability phenomena in the strip are most pronounced between $H_{on}^s$ and $H_{DT}$ lines. $H_{c2}$ is defined resistively at 10\% of the normal state resistance.
Inset: {\em dc} $I_c$ vs. field in the vicinity of the reentrant $H_{DT}$ line at 4.2 K in the strip ($\bigcirc $) and Corbino ($\bullet $) geometries.

Fig. 3. The voltage vs. field at 20 mA at $T$ = 4.3 K (a) and 4.6 K (b) measured with {\em dc} current in the Corbino ($\bullet $) and strip
($\bigcirc $), and with 772 Hz {\em ac} current in the strip (dashed line). $H_{DT}$ marks the position of the thermodynamic phase transition, and $H_{on}^s$ is the onset of an observable {\em dc} response in the strip geometry.



\begin{references}
\bibitem{note1} Note that $\tau_r$ and $L_r$ are also strong functions of the vortex velocity \cite{hend96}. The rapid decrease of $L_r$ with the velocity may account for the reported negative differential resistance below $T_{DT}$ \cite{zhukov}.
\end{references}
\end{document}